\documentclass[aps,prb,twocolumn,showpacs]{revtex4}

\usepackage{amsmath}
\usepackage{amssymb}
\usepackage{epsfig}
\usepackage{enumerate}

\newcommand{\tp}{\tau_\varphi}
\newcommand{\tps}{\tau_{\varphi,S}}
\newcommand{\Lp}{L_\varphi}
\newcommand{\sub}[1]{_{\text{#1}}}
\newcommand{\super}[1]{^{\text{#1}}}
\newcommand{\opn}[1]{\operatorname{#1}}

\begin{document}
\pagestyle{empty}

\title{Magnetic field dependence of dephasing rate due to diluted Kondo impurities}
\author{T. Micklitz,$^1$ T. A. Costi,$^2$ and A. Rosch$^1$ }
\affiliation{$^1$ Institute for Theoretical Physics, University of Cologne, 50937
Cologne, Germany \\
$^2$ Institut f\"ur Festk\"orperforschung, Forschungszentrum J\"ulich, 
52425 J\"ulich, Germany.} 

\date{\today} 
\begin{abstract}
  We investigate the dephasing rate, $1/\tp$, of weakly disordered
  electrons due to scattering from diluted dynamical impurities. Our
  previous result for the weak-localization dephasing rate is
  generalized from diluted Kondo impurities to {\em arbitrary}
  dynamical defects with typical energy transfer larger than $1/\tp$. 
  For magnetic impurities, we study the
  influence of magnetic fields on the dephasing of Aharonov-Bohm
  oscillations and universal conductance fluctuations both analytically
  and using the 
  numerical renormalization group. These results are compared to
  recent experiments.
\end{abstract}
\pacs{72.15.Lh, 72.15.Qm, 72.15.Rn}
\maketitle

\section{Introduction}
Decoherence is the fundamental process leading to a suppression of
quantum mechanical interference and therefore is indispensable for our
understanding of the appearance of the classical world. The
destruction of phase coherence in a quantum system occurs due to
interactions with its environment and can be studied, e.g.,~in
mesoscopic metals and semiconductors where the quantum-mechanical wave
nature of the electrons leads to a variety of novel transport
phenomena at low temperatures.

Although the concrete definition of the dephasing rate, $1/\tp$,
depends on the experiment used to determine it, the electron-electron
interactions are thought to be the dominant mechanism for the
destruction of phase coherence in metals without dynamical impurities
below about 1 K. The dephasing rate for interacting electrons in a
diffusive environment was first calculated by Altshuler, Aronov and
Khmelnitsky (AAK) and vanishes at low temperatures, $T$, with some
power of $T$, depending on the dimensionality of the
system.~\cite{AlArKh82}

In the last decade several independent groups performed
magnetoresistance experiments~\cite{MoJaWe97, ScBaRaSa03, HaVrBr87,
  PiGoAnPoEsBi03} to probe the influence of dephasing on weak
localization in disordered metallic wires. Irritatingly, a saturation
of the dephasing rate, $1/\tp$, has been observed at the lowest
experimentally accessible temperatures. This observation has triggered
an intense discussion on the mechanism responsible for the excess of
dephasing.~\cite{GoZa98, AlAlGe99,De03} The most promising
candidates to explain the saturation of $1/\tp$ are extremely low
concentrations of dynamical impurities, such as atomic two-level 
systems \cite{zawa,imry} or magnetic impurities \cite{mohawebb,ScBaRaSa03,HaVrBr87,PiGoAnPoEsBi03,VaGlLa03,MiAlCoRo06,ZaBoDeAn04}.  
This has been corroborated on the one
hand by experiments\cite{PiGoAnPoEsBi03, ScBaRaSa03} on extremely clean Ag and Au
samples where the dephasing rate continues to decrease well below
100~mK and on the other hand by doping studies with magnetic impurities.~
\cite{MoJaWe97,PiGoAnPoEsBi03,ScBaRaSa03,MaBau06, AlzBir06} As expected
theoretically,~\cite{VaGlLa03} these experiments show a
saturation of $1/\tp$ above the Kondo temperature, $T\sub{K}$, the
characteristic scale of screening of the magnetic moment, and a
suppression of $1/\tp$ below this scale. Recent highly controlled
experiments,~\cite{MaBau06,AlzBir06} in which a few ppm (parts per
million) of Fe ions have been implanted 
by ion beam lithography
into very clean Ag samples, showed that the screening of these Fe ions is
surprisingly well described \cite{MaBau06} by the theoretically
predicted dephasing rate for spin-1/2 Kondo impurities
\cite{MiAlCoRo06} down to temperatures of $0.1 \, T\sub{K}$. At the lowest
temperature, again a plateau in the dephasing rate has been observed
proportional to the number of implanted Fe ions. The origin of this
puzzling behavior is still unclear but may arise from further
dynamical defects created during the implantation process or by rare
Fe ions with a different chemical environment and strongly reduced
magnetic screening.

An obvious option to study the influence of magnetic impurities on the
dephasing rate is to measure its dependence on an externally applied
magnetic field. The application of sufficiently large magnetic fields
freezes out inelastic spin-flip processes as discussed theoretically
in Refs.~[\onlinecite{VaGlLa03}], and therefore one expects
the dephasing rate to return to the value predicted by AAK for
dephasing induced by Coulomb interactions in a diffusive environment.
The orbital contribution of the magnetic field does, however, destroy
the weak-localization (WL) contribution to the magneto-resistance, as
the joint propagation of an electron and a hole along {\em
  time-reversed} trajectories (the Cooperon) picks up extra (random)
Aharonov-Bohm phases in the presence of  external magnetic fields.
Measuring the $B$-dependent dephasing rate in a WL experiment is
therefore only possible in strictly one- or two-dimensional systems
using magnetic fields almost exactly parallel to such a structure,
requiring an accurate alignment of magnetic fields.

Universal conductance fluctuations (UCF) and Aharonov-Bohm (AB)
oscillations with a periodicity of $h/e$, on the other hand, are not suppressed by orbital effects
and can be used rather directly to determine the field dependence of
the dephasing rate.~\cite{h2e} 

UCFs can be observed as characteristic
fluctuations of the conductance as a function of the magnetic field. The
external magnetic field enters the metal and changes the pattern of
the electrons wave functions and therefore the conductance in a random
but reproducible way (``magnetofingerprint"). In AB experiments
performed on mesoscopic rings, these sample fluctuations are further
modulated by periodic $h/e$-oscillations resulting from the magnetic
flux piercing the ring. Both UCF and AB oscillations rely on the
constructive interference occurring in the collective propagation of
electrons and holes traveling along the {\em same} path (the
diffuson).  These are robust against the breaking of time-reversal
invariance (while Cooperon contributions are rapidly suppressed by
small fields) but are sensitive to dephasing by inelastic processes.
Indeed,  Benoit {\it et al.} \cite{benoit} and Pierre {\it et al.}  \cite{PiGoAnPoEsBi03} have shown that the
amplitude of Aharonov-Bohm oscillations {\em increases} by almost an
order of magnitude for increasing magnetic fields clearly showing a
suppression of dephasing by magnetic (Zeeman) fields. This leads to the
conclusion that the main mechanism of dephasing in the investigated
low-temperature regime is the scattering from magnetic impurities.

Previously we have studied the zero-field dephasing rate due to
diluted Kondo impurities as measured from the WL experiment.~\cite{MiAlCoRo06} 
We showed that the dephasing rate for all
experimentally relevant temperatures is proportional to the inelastic
crosssection,~\cite{ZaBoDeAn04} which itself can be expressed in
terms of the $T$-matrix describing the scattering of the electrons
from a single magnetic impurity. Such a relation has been proposed previously by Schwab and
Eckern \cite{schwabeckern} in the context of UCFs.

In this paper we generalize our previous work to arbitrary, diluted
impurity scatterers with typical energy transfer larger than $1/\tp$. 
Furthermore we supply the magnetic field
dependence of $1/\tp$ as measured in the AB experiment (and compare
our results to the AB experiments performed by Pierre and Birge.~\cite{PieBir02}) 
The outline of this paper is as follows: In Sec.~II
we discuss the dephasing rate due to generic dynamical scatterers as
measured in the WL experiment. We briefly review our previous results \cite{MiAlCoRo06} and generalize them to arbitrary dynamical impurities with typical energy transfer larger than $1/\tp$.
In Sec.~III we turn to
the dephasing rate as measured from the UCF and the amplitude of the
AB oscillations. We briefly review the main concepts entering the
analysis of these experiments and discuss how the dephasing rate
measured from the UCF differs from that measured in the WL experiment.
The main goal of Sec.~III is to give the dephasing rate as measured
from the amplitude of the AB oscillations.  Results for the dephasing
rate obtained using the numerical renormalization group (NRG) are
described in Sec.~IV. Sec.~V summarizes with a discussion.

\section{Dephasing rate from weak-localization corrections to the conductivity}

\begin{figure}[b]
\centering \includegraphics[width=4cm]{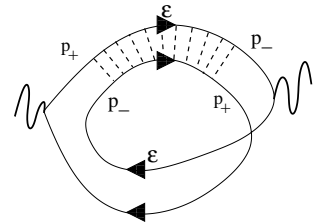} 
\caption{\label{fig2}Diagrammatic representation of the Cooperon $C^0_{\Omega=0}(q)$ which enters the WL corrections to the Drude conductivity. $p_\pm=q/2\pm p$ and wavy lines denote current operators.}
\end{figure}

\begin{figure}[t]
\centering \includegraphics[width=7.4cm]{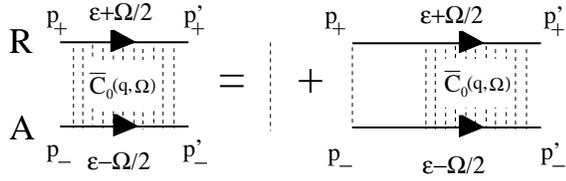}
\caption{\label{barecoop} Bethe-Salpeter equation for the bare Cooperon $\bar{C_0}(q, \Omega)=\frac{1}{2\pi\nu\tau^2}C_0(q, \Omega)$, $p_\pm=q/2\pm p$, and $p'_\pm=q/2\pm p'$. Dashed lines denote scattering from static impurities and $R$, $A$ denotes the particle and hole lines, respectively.}
\end{figure}

The most accurate way to extract $1/\tp$ at low magnetic fields is via the WL corrections to the Drude conductivity, which result from coherent back scattering of an electron-hole pair traveling along time-reversed paths in the disordered environment. Technically, the coherent propagation of the electron-hole pair is described by the Cooperon, $C_\Omega(\bold{q})$, and the WL correction is given by
\begin{align}\label{wl}
\Delta \sigma_{\rm WL}^0= -\frac{2e^2D}{\pi} \int \frac{d^d \bold{q}}{(2\pi)^d} C_{\Omega=0} (\bold{q}),
\end{align} see Fig. \ref{fig2}. In the absence of dephasing by inelastic processes $C_\Omega(\bold{q})$ is the bare Cooperon,
\begin{align}
C^{0}_{\Omega} (\bold{q})
=  \frac{1}{D\bold{q}^2+i\Omega + 1/\tau_B},
\end{align} as diagrammatically depicted in Fig.~\ref{barecoop} and\cite{AlAr85} 

\begin{align}
\label{tb}
&\frac{1}{\tau_B} = 4DeB, \quad d=2 \quad(\bold{n}~ ||~ \bold{B}), \\
&\frac{1}{\tau_B} = \frac{D}{3}(eBL_\perp)^2, \quad d=1,2 \quad(\bold{n} \perp\bold{B}),
\end{align} is the dephasing rate due to the applied magnetic field, $B$. Here $D$ is the diffusion constant, $d$ is the dimension of the diffusion process, $\tau$ denotes the mean scattering time corresponding to a mean free path $l=v\sub{F}\tau$, $\bold{n}$ is a unit vector orthogonal to the probe in $d=2$ and pointing along the wire in $d=1$ and $L_\perp$ is the transverse dimension of the sample. As can be seen from Eq.~(\ref{wl})
the WL corrections depend on the strength of the applied magnetic field, $B$,  and diverge in low dimensions, $d=1, 2$ for $B=0$, reflecting the fact that WL corrections in low-dimensional systems may become strong and lead to strong Anderson localization.
 
Taking into account interactions (as e.g. provided by dynamical impurities) the bare Cooperon dresses with a mass, $1/\tp$, i.e. (if purely exponential decay is guaranteed),
\begin{align}
C_{\Omega=0} (\bold{q})
= \frac{1}{D\bold{q}^2+1/ \tp +1/\tau_B}.
\end{align} For weak magnetic fields ($\tp \ll \tau_B$) the WL corrections are therefore cut off by $1/\tp$, allowing to determine the dephasing rate from fitting the magnetoresistance. 

In this section we study the dephasing rate due to {\em generic}
diluted, dynamical scatterers with typical energy transfer larger than $1/\tp$ 
as measured from WL. To be specific, we
consider a Hamiltonian of the general form
\begin{align}
\label{himp}
H\sub{imp}=H\sub{imp}^0+\sum_i c^\dagger_{\bold{k}\sigma}c_{\bold{k}'\sigma'} f_{\bold{k}\bold{k'}\sigma\sigma'}^\alpha \hat{X}_\alpha e^{i(\bold{k}-\bold{k}')\bold{x}_i},
\end{align} where $H\sub{imp}^0$ is the Hamiltonian of the isolated
impurity, $c^\dagger, c$ are creation and annihilation
operators of conduction band electrons, $\bold{x}_i$ denotes the
position of the impurities, and the momentum and spin dependent
function $f^\alpha$ parametrizes the coupling to some operator
$\hat{X}_\alpha$ describing transitions of the internal states of the
dynamical impurity. Eq.~(\ref{himp}), e.g., describes the coupling of
the conduction band electrons to Kondo impurities, two-level systems,
etc.  In order to find the dephasing rate due to such generic
impurities one has to compute the ``self energy" or ``mass" of the
Cooperon generated by the operators of Eq.~(\ref{himp}).

\begin{figure}[t]
\centering \includegraphics[width=8.2cm]{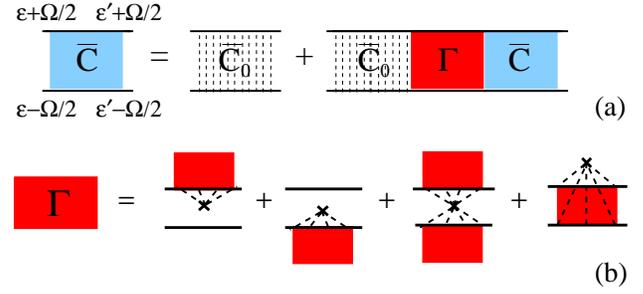}
\caption{\label{fig1} Bethe-Salpeter equation for the Cooperon $\bar{C}$ in
  the presence of dilute dynamical impurities to linear order in $n_i$. $\bar{C}_0$ is the bare
  Cooperon in the absence of interactions and $\Gamma$ the irreducible
  vertex obtained by adding self-energy, elastic-, and inelastic-vertex
  contributions. The crosses with attached dashed lines denote the averaging
over impurity positions $\bold{x}_i$, the squares the inelastic scattering from a single impurity to arbitrary order.}
\end{figure}

Two assumptions allow to reduce this problem to that of summing up a
simple geometric series.  First, we assume that the concentration of
dynamical impurities $n_i$ is small, and, second, that the elastic
mean free path $l$ is large. How small $n_i$ and how large $l$ has to be, depends on both,
the dynamics and the extension of the impurity as briefly discussed below and -- for Kondo impurities --
in more detail in Ref.÷[\onlinecite{MiAlCoRo06}] (the influence of
stronger disorder on the distribution of Kondo temperatures has
recently been investigated by Kettemann and Mucciolo\cite{kettemann}).
Note that in relevant experimental systems in the WL regime \cite{MoJaWe97, ScBaRaSa03, HaVrBr87,
  PiGoAnPoEsBi03} the two assumptions are well
justified÷\cite{MiAlCoRo06}. The first observation is that quantum interference corrections to the inelastic scattering rate
are small when a diffusing electron is unlikely to return to the same dynamical impurity (i.e., when the weak localization corrections are weak). Technically speaking, this is reflected in the fact that diagrams mixing scattering 
from dynamical and static impurities are suppressed by factors  of $1/N_\perp$ and $a/l$ [or $1/(k_F l)$ for
$a<1/k_F$], where  $N_\perp$ is the number of transverse channels in a quasi 1 or 2 dimensional system and $a$ is the typical diameter of the dynamical impurity.
This effect and further system dependent factors relevant for the suppression of quantum interference corrections are discussed in Ref.÷[\onlinecite{MiAlCoRo06}].
Only at lowest, experimentally unprobed temperatures, does
the enhanced infrared singularity, caused by the presence of extra
diffusion modes, overcompensate this phase space suppression factor,
as discussed in detail in Ref.~[\onlinecite{MiAlCoRo06}]. The small parameter
$1/(k\sub{F} l)$ or $a/l$ therefore reduces the problem to compute the
'mass' of the Cooperon to that of solving the Bethe-Salpeter equation
diagrammatically depicted in Fig.~\ref{fig1}(a).  

\begin{figure}[b]
\centering \includegraphics[width=3.5cm]{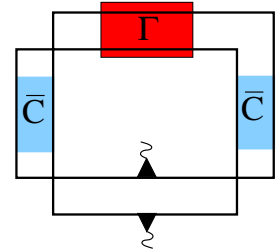}
\caption{\label{wlcorrections} Lowest order correction to WL contributions due to inelastic vertex. The Cooperon, $\bar{C}$, is dressed with a mass resulting from summation of the elastic part of the vertex $\Gamma$, i.e., the self energy and the elastic vertex contribution.}
\end{figure}

For small $n_i$, one
can furthermore restrict the analysis of the irreducible vertex
$\Gamma$ to terms linear in $n_i$ as shown in Fig.~\ref{fig1}(b).
$\Gamma$ can be separated into three distinct contributions:
self-energy diagrams [the first two terms in Fig.~\ref{fig1}(b)], an
`elastic' vertex correction with no energy transfer between upper and
lower line (third term), and an `inelastic' vertex where interaction
lines connect the two lines (last term). Only this inelastic vertex makes
the Bethe-Salpeter equation a true integral equation as it mixes
frequencies but, fortunately, this term can be neglected~\cite{AlArKh82} if the
typical energy, $\Delta E$, exchanged between electrons and holes
during an interaction process greatly exceeds the dephasing rate due
to the dynamical impurities, $1/\tp$, i.e., $\Delta E\tp\gg1$. 
Physically,~\cite{AlArKh82} the suppression of the inelastic vertex arises as an
exchange of  energy $\Delta E$ leads to a phase mis-match of order $e^{i
  \Delta E \tp}$ between electron and hole destroying interference
  completely for $\Delta E\tp\gg
1$. Technically, one can confirm this argument by estimating corrections to
the WL contributions due to the inelastic vertex, as, e.g., depicted in
Fig.~\ref{wlcorrections}.~\cite{MiAlCoRo06} More importantly, however, the condition $\Delta E\tp\gg1$ always holds for sufficiently
small concentrations $n_i$, since $1/\tp \propto n_i$. In the case of Kondo impurities discussed below, for example, this condition translates to $n_i\ll\nu T\sub{K}$.

Restricting to the self-energy and elastic vertex contributions, the Bethe-Salpeter equation is easily solved: Since self-energy and elastic vertex contributions conserve the energy
of single electron lines, the solution of the reduced Bethe-Salpeter equation
amounts to a straightforward summation of a geometric series.  Setting
the center-of-mass frequency $\Omega$ to $0$ (see Fig.~\ref{fig2}, and Fig.~\ref{fig1}),
the Cooperon is given by 

\begin{align}
C_{\Omega=0} (\epsilon, \bold{q})
= \frac{1}{D\bold{q}^2+1/ \tp(\epsilon,T) +1/\tau_B},
\end{align} with the $T$ and $\epsilon$ dependent dephasing rate
\begin{align}
\label{gen1}
\frac{1}{\tp(\epsilon,T)} &= \frac{2n_i}{\pi\nu} \Bigg[ \int \frac{d^3\bold{p}}{(2\pi)^3} g_{\epsilon}(\bold{p}) \frac{1}{2i} \left[  T\super{A}_{-\bold{p},-\bold{p}}(\epsilon) - T\super{R}_{\bold{p}\bold{p}}(\epsilon) \right] \nonumber \\
- \int& \frac{d^3\bold{p}}{(2\pi)^3} \int \frac{d^3\bold{p}'}{(2\pi)^3} g_{\epsilon}(\bold{p})g_{\epsilon}(\bold{p}')T\super{R}_{\bold{p}\bold{p}'}(\epsilon) T\super{A}_{-\bold{p},-\bold{p}'}(\epsilon) \Bigg].
\end{align} Here $g_{\epsilon}(\bold{p})=\frac{\pi/2\tau
}{[\epsilon(\bold{p})-\epsilon]^2+\frac{1}{4\tau^2}}$ restricts the
electrons momenta, $\bold{p}$, and energies, $\epsilon$, to the
Fermi-surface, $\epsilon(\bold{p})$ is the dispersion relation of the
conduction band, $ T\super{A,R}$ are the advanced/retarded T-matrices, defined by the Green function $G_{\bold{x}\bold{x}'}(\epsilon)=G^0_{\bold{x}\bold{x}'}(\epsilon) + G^0_{\bold{x}\bold{0}}(\epsilon)T(\epsilon)G^0_{\bold{0}\bold{x}'}(\epsilon)$, 
and $\nu$ denotes the density of states per spin. Eq.~(\ref{gen1})
generalizes our result of Ref.~[\onlinecite{MiAlCoRo06}] to arbitrary shaped
diluted impurities. Notice that also forward scattering processes
enter $1/\tp$, which do not contribute to the transport scattering rate. We stress that Eq.~(\ref{gen1}) is the general result for the
dephasing rate for a weakly disordered metal due to a low
concentration of {\em generic} dynamical impurities for which the
condition $\Delta E\gg1/\tp$ holds. In the opposite limit, $\Delta E \gg1/\tp$, vertex corrections become important, as has been discussed earlier on \cite{HiLaNa80,falko} in the context of magnetic impurities.

As we assumed that $a \ll l$, 
Eq.~(\ref{gen1}) can be further simplified,
\begin{align}
\label{gen2}
&\frac{1}{\tp(\epsilon,T)} = \frac{2n_i}{\pi\nu} \Bigg[ \int_{S_F^\epsilon} \frac{d^2\bold{p}}{(2\pi)^3}  
\frac{1}{|v\sub{F}(\bold{p})|} \opn{Im}\left[\pi T\super{A}_{\bold{p}\bold{p}}(\epsilon)\right] \nonumber \\
&\quad-\int_{S_F^\epsilon} \frac{d^2\bold{p}}{(2\pi)^3} \int_{S_F^\epsilon} \frac{d^2\bold{p}'}{(2\pi)^3} \frac{1}{|v\sub{F}(\bold{p})|} \frac{1}{|v\sub{F}(\bold{p}')|} |\pi T\super{R}_{\bold{p}\bold{p}'}(\epsilon)|^2 
\Bigg],
\end{align} where $S_F^\epsilon$ is the Fermi-surface (or more
precisely the surface with $\epsilon_k=\epsilon$). Here we also
assumed a time-reversal invariant system with 
$T\super{R}_{\bold{p}\bold{p}'}(\epsilon)=T\super{R}_{-\bold{p}',-\bold{p}}(\epsilon)$
and employed the identity
$\left[T\super{R}_{\bold{p}\bold{p}'}(\epsilon)\right]^*=T\super{A}_{\bold{p}'\bold{p}}(\epsilon)$.
Also Kettemann and Mucciolo have
generalized the dephasing rate for a momentum independent $T$ matrix to
Eq.~(\ref{gen2})  independently in a recent report.~\cite{kettemann}
The dephasing rate given in Eq.~(\ref{gen2}) has a simple
interpretation \cite{ZaBoDeAn04}: Since the Fermi-surface
integrated imaginary part of the $T$-matrix is proportional to the
total cross-section and $|T\super{R}_{\bold{p}\bold{p}'}|^2$ (integrated
over the Fermi surface) is proportional to the elastic cross section,
its difference is, by definition, proportional to the {\em inelastic}
cross section, $\sigma_{\rm inel}$, introduced in
Ref.~[\onlinecite{ZaBoDeAn04}]. Therefore, Eq.~(\ref{gen2}) can be
rewritten in the form
\begin{align}\label{semi}
\frac{1}{\tp(\epsilon)}= n_i \left\langle\,  v_F(\bf {p}) \, \sigma_{\rm inel}(\bf p,\epsilon)\, \right \rangle, 
\end{align} 
where $\langle ...  \rangle $ denotes an angular average weighted by
$1/v_F(\bf {p})$ to take into account that fast electrons are
scattered more frequently from elastic impurities. According to Eq.
(\ref{semi}), $\tp$ is nothing but the average time needed (in a
semiclassical picture) to scatter from an impurity with cross section
$\sigma_{\rm inel}$.  Note that the vanishing of $1/\tp$ for static
impurities is guaranteed by the optical theorem.

From Eq.~(\ref{gen2}) we can read off the dephasing rate for diluted
dynamical isotropic s-wave scatterers\cite{MiAlCoRo06, schwabeckern}
\begin{align}\label{tau}
\frac{1}{\tp(\epsilon,T)} = \frac{2n\sub{i}}{\pi\nu} \left[
  \pi\nu^{\rm loc}
  \opn{Im}\!\!\left[ \opn{T}\super{A}(\epsilon) \right] - |\pi\nu^{\rm loc}
  \opn{T}\super{R}(\epsilon)|^2 \right],
\end{align}
where $\nu^{\rm loc}$ is the {\em local} density of states at the
Fermi energy at the site of the impurity which can differ from the
thermodynamic density of states entering the prefactor. Note that in
the case of Kondo impurities discussed below (and 
in Ref.~[\onlinecite{MiAlCoRo06}]), the combination $\nu^{\rm
  loc} T^{R/A}(\epsilon)=f(\epsilon/T\sub{K},T/T\sub{K},B/T\sub{K})$ is an {\em
  universal} dimensionless function of the ratios
$\epsilon/T\sub{K},T/T\sub{K},B/T\sub{K}$. If the assumptions underlying the
derivation of Eq.~(\ref{tau}) are valid, one can therefore predict
{\em without} any free parameter the dephasing rate if the
concentration of spin-1/2 impurities, the Kondo temperature and the
thermodynamic density of states are known (see, e.g.,
Ref.~[\onlinecite{MaBau06}]). However, one of the assumptions
underlying the derivation of the prefactor of Eq.~(\ref{tau}) may not
be valid in realistic materials: we assumed that the static impurities
are completely uncorrelated and local such that electrons are
scattered uniformly over the Fermi surface. While this should be a good
assumption in doped semiconductors, this may not be valid in metals
with complex Fermi surfaces and strongly varying Fermi
velocities. Under the latter conditions, we expect that the prefactor
of Eq.~(\ref{tau}) becomes nonuniversal, yielding temperature-independent
  corrections of order one, which may be important for the
  interpretation of high-precision experiments.~\cite{AlzBir06, MaBau06}

In Ref.~[\onlinecite{MiAlCoRo06}] we have calculated the leading
corrections to Eq.~(\ref{tau}) arising from mixed diagrams involving
combined scattering from static and dynamical impurities and from
diagrams including higher processes in $n_i$. We showed that,
suppressed by the small parameter $1/(k\sub{F}l)$, their contributions
are negligible at all experimentally relevant temperatures, $T$. Only
at the lowest experimentally irrelevant temperatures do these 
corrections become important due to infrared singularities 
of the dressed interaction potential (dressed by coherent
backscattering processes). The estimates
of subleading corrections presented in Ref.~[\onlinecite{MiAlCoRo06}]
can be generalized to extended dynamical impurities by replacing
$1/(k_F l)$ by $a/l$ for $k_F a >1$.

The experimentally measured dephasing rate does not resolve the
dependence on the electrons energy $\epsilon$. We described in
Ref.~[\onlinecite{MiAlCoRo06}] that to allow for a comparison with the
$\epsilon$--independent dephasing rate, $\tp^{-1}(T)$, extracted from
the WL experiment, the energy--resolved representations of $1/\tp$,
Eq.~(\ref{gen2})-(\ref{tau}), still require an average over energies
according to

\begin{align}
\label{av}
\frac{1}{\tp(T)} =\left\{
\begin{array}{ll}
 \left[ - \int d\epsilon f\sub{F}'(\epsilon)
\tp(\epsilon,T)^{\frac{2-d}{2}} \right]^{\frac{2}{d-2}} & d=1,3, \\[2mm]
\frac{1}{\tau}\exp\!\left[ \int d\epsilon f\sub{F}'(\epsilon)
\ln \frac{\tp(\epsilon,T)}{\tau}\right] & d=2,\\[2mm]
 - \int d\epsilon f\sub{F}'(\epsilon)
/\tp(\epsilon,T) & \tp/\tau_B \gg 1. \\
\end{array}\right.
\end{align}  Here the last line applies  to a case where a relatively strong magnetic field, $B$,
is present.

Specifying to a situation where the coupling of the conduction band electrons to the (diluted) dynamical impurities is described by the spin-1/2 Kondo effect, the general Hamiltonian of Eq.~(\ref{himp}) takes the form

\begin{align}\label{h_s}
H\sub{S} = J \sum_i \hat{\bold{S}}_i c^\dagger_\sigma(\bold{x}_i)  \bold{\sigma}_{\sigma\sigma'} c_{\sigma'}(\bold{x}_i),
\end{align} where $J$ is the exchange constant. An external magnetic field, $B$, causes Zeeman splitting, $\epsilon_z$, of the conduction band electron spin states,

\begin{align}
\label{ze}
\epsilon_z = g_e\mu_B B,
\end{align} and couples to the impurity spins according to

\begin{align}
\label{sb}
H_B = g_S\mu_B B\sum_i \hat{\bold{S}}^z_i. 
\end{align} $g_e$, $g_S$ are the electrons and the magnetic impurities
gyromagnetic factors, respectively. As already mentioned in the
introduction, measuring the $B$-field dependence of the dephasing rate
due to Kondo impurities in a WL experiment is a highly delicate task.
 In view of this
difficulty it is more feasible to measure the $B$-dependence of
$1/\tp$ from the amplitude of 
the AB oscillations as discussed in the
following section.

\section{Dephasing rate from universal conductance fluctuations and Aharanov-Bohm oscillations}

Let us first begin with a brief discussion of the universal conductance fluctuations (UCF) and their dependence on $1/\tp$ and then turn to the experiment on Aharonov-Bohm rings.  
 
To be specific, we consider a wire of non-interacting electrons,
scattering elastically from static impurities, and inelastically off a
low concentration, $n_i$, of Kondo impurities, where the coupling
of the conduction band electrons to the dynamical impurities is
described by the Hamiltonian Eq.~(\ref{h_s}) and the influence of the
magnetic field is accounted for by Eqs.~(\ref{ze}) and (\ref{sb}).

\begin{figure}[b]
\centering \includegraphics[width=7.4cm]{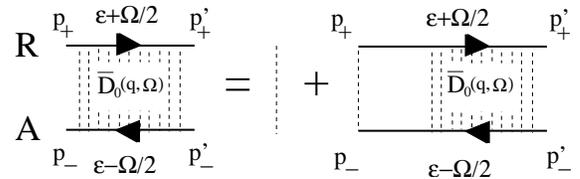}
\caption{\label{barediff} Bethe-Salpeter equation for the bare diffuson $\bar{D}_0(q, \Omega)=\frac{1}{2\pi\nu\tau^2}D_0(q, \Omega)$. $p_\pm=p\pm q/2$ and $p'_\pm=p'\pm q/2$. Dashed lines denote scattering from static impurities and $R$, $A$ denotes the particle and hole lines, respectively.}
\end{figure}
For wires of length $L\gg L_T=\sqrt{D/T}$, the fluctuations of the
 conductance, $\overline{\delta g\delta g}$, are
determined\cite{AronovSharvin87} by 
\begin{align}
\label{D}
\overline{\delta g\delta g} = \frac{(2e^2D)^2}{3\pi TL^4}& \int d\epsilon_1 d\epsilon_2 f'\sub{F}(\epsilon_1)f'\sub{F}(\epsilon_2) \nonumber\\
& \int d\bold{x}_1 d\bold{x}_2|P_{\epsilon_1,\epsilon_2}(\bold{x}_1,\bold{x}_2)|^2.
\end{align} Here $P_{\epsilon_1,\epsilon_2}(\bold{x}_1,\bold{x}_2)$ is
the amplitude for an electron-hole pair, with energies $\epsilon_1,
\epsilon_2$ respectively, to diffusively travel from $\bold{x}_1$ to
$\bold{x}_2$ along the same trajectory (diffuson, see
Fig.~\ref{barediff}). The overbar denotes the ensemble average, which
is experimentally realized by changing the magnetic field.
Eq.~(\ref{D}) assumes the large ring diameters, $L \gg \Lp$, such that the dephasing rate,
$1/\tp$, controls the magnitude of the
fluctuations. Furthermore we assume temperatures $T\tp\gg1$. Notice that generally there is also a
contribution from the Cooperon, which is, however, suppressed already
for small magnetic fields. 

\begin{figure}[t]
\centering \includegraphics[height=3cm,width=6cm]{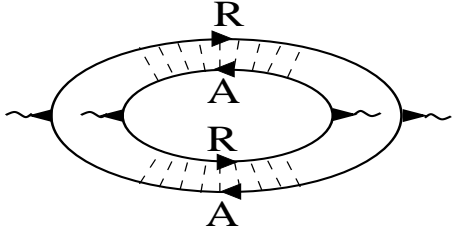}
\caption{\label{ucf-d} Diagram giving the main contribution to the UCF. Dashed lines represent coherent impurity scattering of electron (R) hole (A) pair, i.e. the bare diffuson where interaction due to scattering from magnetic impurities is not yet taken into account. }
\end{figure}
\begin{figure}[t] 
\centering \includegraphics[width=8.2cm]{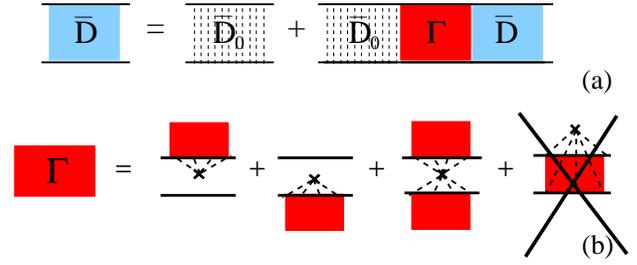}
\caption{\label{diff} (a) Bethe-Salpeter equation for the diffuson, $\bar{D}$, in
  the presence of (dilute) magnetic impurities to linear order in $n\sub{i}$. $\bar{D}_0$ is the bare
  diffuson in the absence of interactions. (b) Diagrammatic representation of the irreducible
 interaction vertex,  $\Gamma$,  consisting of the self-energy (represented by the first two contributions), the elastic vertex (third contribution).  The inelastic vertex (the fourth contribution) does not enter the diffuson as measured in the UCF.}
\end{figure}

\begin{figure}[b]
\centering \includegraphics[height=2cm,width=6cm]{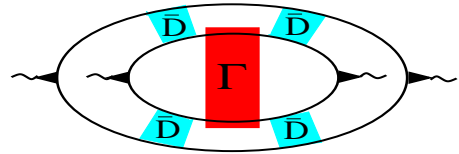}
\caption{\label{ducf-d}Diagrammatic representation of lowest order corrections to UCF due to inelastic vertex contributions. Notice that for a local interaction these are not only suppressed by powers of 
$1/(\Delta E \tp)$ but also small in powers $1/(k\sub{F} l)$.}
\end{figure}

 
We change to momentum representation and separate the two-particle
propagator $P$ into its spin-singlet and -triplet components, $P=
\sum_{i=1,..,4} P^{(i)}$, where (following the notation of Ref.~\onlinecite{VaGlLa03})
\begin{align}
\label{st}
&P^{(i)}_{\epsilon_1\epsilon_2}(\bold{q}) = \nonumber \\
&\quad  \frac{1}{D\bold{q}^2 + i(\epsilon_1-\epsilon_2+\zeta_i\epsilon_z)+1/\tau^{(i)}_{SO}+1/\tps^{(i)}(\epsilon_1,\epsilon_2,B)}. 
\end{align} The modes $i=1,2,3$ describe the spin triplet state with
$S_z$ component equal to $1,-1$, and $0$, respectively. $i=4$ denotes
the spin singlet channel. The Zeeman splitting enters only the
triplet-diffuson with nonvanishing projection $S_z=\pm 1$, i.e.
$\zeta_i=\pm1$ for $i=1,2$ and zero otherwise. $1/\tau^{(i)}\sub{SO}$
is the spin-orbit scattering rate. $1/\tau^{(i)}\sub{SO}$ is identical
for the three spin triplet-diffuson ($i=1,2,3$) and zero for the spin
singlet mode ($i=4$).  For strong spin-orbit scattering only the
singlet diffusion contributes (otherwise $1/\tau^{(i)}\sub{SO}$ is an
additional fitting parameter).  Finally $1/\tps^{(i)}$ is the
dephasing rate for the $i$th diffuson mode due to the presence of
diluted magnetic impurities which has the structure
\begin{align}
\label{tucf}
\frac{1}{\tps^{(i)}(\epsilon_1,\epsilon_2)} = &\frac{2n_i}{\pi\nu} \Bigg(
\frac{ \pi\nu}{2i} \big[  \opn{T}^{(i,a)}(\epsilon_2,B) - \opn{T}^{(i,b)}(\epsilon_1,B) \big] 
 \nonumber \\ 
& - (\pi\nu)^2\opn{T}^{(i,c)}(\epsilon_1,B)\opn{T}^{(i,d)}(\epsilon_2,B)\Bigg),
\end{align} 
where the proper combination of $T$-matrices for the
various channels can be read off by comparison with Table~\ref{tmatrices}.
Eq.~(\ref{tucf}) is evaluated from summing up self-energy and elastic
vertex contributions. Notice that in contrast to the WL experiment the
electron and hole lines (i.e., the inner and outer rings) in
Fig.~\ref{ucf-d} represent different measurements. Therefore there are
no correlations between dynamical impurities residing on different
rings and interaction lines may only be drawn within the same ring.
Consequently the inelastic vertex contributions do not enter the
Bethe-Salpeter equation for the diffuson, see Fig.~\ref{diff}.
Notice that there are inelastic vertex contributions, as, e.g., depicted in 
Fig.~\ref{ducf-d}, which become important in the context of
electron-electron interactions.~\cite{AleinerBlanter} It is instructive to compare those to the inelastic vertex corrections relevant for WL depicted in Fig. \ref{wlcorrections}. In the latter case, the sum of the incoming momenta of the vertex is small due to the Cooperon in Fig. \ref{wlcorrections}. Consequently, the inelastic vertex corrections to the WL dephasing rate are {\em not} suppressed by powers of $1/(k\sub{F} l)$ but only by powers of $1/(\Delta E \tp)$. In contrast, the relevant momenta in 
Fig.~\ref{ducf-d} are uncorrelated (i.e., particle and hole are far apart), leading to an 
suppression both by powers of $1/(k\sub{F} l)$ and of $1/(\Delta E \tp)$.~\cite{mirlinring}

\begin{table}[h]
\centering
\begin{tabular}{|p{.31cm}|p{2.2cm}|p{5.9cm}|}
\hline
\hline
$i$ & $|S,M\rangle$ & Combinations of $T$-matrices\\
\hline
1 & $S=1, M=1$ & $T^1 = \frac{1}{2i} \left(T\super{A}_\downarrow - T\super{R}_\uparrow \right)  - T\super{R}_\downarrow T\super{A}_\uparrow$\\
\hline
2 & $S=1, M=-1$ & $T^2 = \frac{1}{2i} \left(T\super{A}_\uparrow - T\super{R}_\downarrow \right)  - T\super{R}_\uparrow T\super{A}_\downarrow$\\
\hline
3 & $S=1, M=0$ & $T^3 = \frac{1}{2}\opn{Im} \left( T\super{A}_\uparrow + T\super{A}_\downarrow \right) - \frac{1}{2} \left(T\super{R}_\uparrow T\super{A}_\uparrow - T\super{R}_\downarrow T\super{A}_\downarrow \right)$\\
\hline
4 & $S=0$ & $T^4 = \frac{1}{2}\opn{Im} \left( T\super{A}_\uparrow + T\super{A}_\downarrow \right) - \frac{1}{2}\left( T\super{R}_\uparrow T\super{A}_\uparrow - T\super{R}_\downarrow T\super{A}_\downarrow \right)$\\
\hline
\hline
\end{tabular}
\caption{Combination of $T$-matrices entering the dephasing rates for spin-triplet and spin-singlet diffusons. $S$ denotes the total spin and $M$ its $z$ component. $T_\uparrow, T_\downarrow$ denotes the $T$-matrix for spin-up and spin-down electrons, respectively.}\label{tmatrices}
\end{table}

We point out the following differences for $1/\tps$ measured from the UCF experiment, Eq.~(\ref{tucf}), compared to that found from the WL, Eq.~(\ref{tau}). First, the $T$-matrices entering Eq.~(\ref{tucf}) depend on the spin configuration of the diffuson-mode and have acquired a $B$-dependence due to the coupling of the impurity spin to $B$, Eq.~(\ref{sb}). Second, $1/\tps$ depends on two energies. This results from the fact, that in the UCF experiment electron and hole lines constituting the diffuson are produced in different measurements of the conductance (see Fig.~\ref{ucf-d}). Therefore their energies are individually averaged as can be seen in Eq.~(\ref{D}).


From Eq.~(\ref{D}) and (\ref{st}) the amplitude of the UCFs is obtained
to be proportional to $\sqrt{\tps}$. Especially compared to the AB
oscillations, discussed below, the  $1/\tps$ dependence of the UCFs is 
 rather weak. In the following, we will therefore focus our discussion
 on AB experiments.

Aharonov-Bohm oscillations are measured in a ring geometry,~\cite{ab1, PieBir02} where the
conductance oscillates periodically as a function of $B$ piercing the
ring, and the amplitude of these AB-oscillations decrease
exponentially with $1/\tps$. The periodic oscillations result from the
change of boundary conditions, due to flux lines piercing the ring and
can be calculated from
\begin{align}
\label{AB}
\overline{\delta g(\phi)\delta g(\phi+\Delta\phi)} = \frac{(2e^2D)^2}{3\pi TL^4}& \int d\epsilon_1 d\epsilon_2 f'\sub{F}(\epsilon_1)f'\sub{F}(\epsilon_2) \nonumber\\
& \int d\bold{x}_1 d\bold{x}_2|P^{\Delta\phi} _{\epsilon_1,\epsilon_2}(\bold{x}_1,\bold{x}_2)|^2.
\end{align} 
Here $P^{\Delta\phi}$ is again given by  Eq.~(\ref{st}) but now the
continuous $\bf q$ have to be replaced \cite{AronovSharvin87} by  discrete
momenta, $\bold{q}= \bold{q}_m(\Delta\phi) = \frac{2\pi}{L}( m +
\frac{\Delta\phi}{\phi_0})$, depending on the difference of the
magnetic flux during the individual measurements of $g$. The
fluctuations are a periodic function in $\Delta\phi/\phi_0$, where
$\phi_0 = 2\pi/e$ is the elementary flux-quantum and
$\Delta\phi=\Delta BL^2/(4\pi)$. Therefore an expansion in its
harmonics can be made,~\cite{AronovSharvin87}

\begin{align}
\label{ak}
\overline{\delta g(\phi)\delta g(\phi+\Delta\phi)} = \frac{C e^4}{\pi^2} \sum_{k=0}^{\infty}{\cal A}_k(B)\cos\left[2\pi k\frac{\Delta\phi}{\phi_0}\right], 
\end{align} where $C$ is a factor of order 1, depending on the sample geometry in the vicinity of the ring.
Restricting to the situation of strong spin-orbit scattering~\cite{SO} where the spin singlet-diffuson gives the leading contributions to Eq.~(\ref{AB}), one finds that (for $L\gg\Lp$)
\begin{align}
\label{a_k}
{\cal A}_k(\bold{B}) = \frac{(2\pi)^3D^{3/2}}{T^2L^3} \int d\epsilon \frac{e^{-\frac{k L}{  \sqrt{ D \tp (\epsilon) }}}}{\cosh^4(\epsilon/2T)} \sqrt{\tp(\epsilon)},
\end{align}  where [$\tp=\tps^{(4)}$ in the notation of Eq.~(\ref{tucf})]

\begin{align}
\label{tab}
&\frac{1}{\tp(\epsilon,T,B)}= \nonumber \\
 &\hspace{1cm} \frac{2n_i}{\pi\nu} \Bigg(
\pi\nu\opn{Im}\left[\opn{T}_{(4)}\super{A}(\epsilon,B)\right]  - |(\pi\nu)\opn{T}_{(4)}\super{R}(\epsilon,B)|^2 \Bigg).
\end{align} Here $\epsilon=\epsilon_1+\epsilon_2$ and we used that
relevant contributions to the integral over energy differences,
$\bar{\epsilon}=\epsilon_1 -\epsilon_2$, result from energies $\bar{\epsilon}
\ll 1/\tp$ to eliminate the $\bar{\epsilon}$ dependence. Notice that such a reduction to a single
energy-integral can only be done in a one-dimensional system where the
$\bold{q}$-integral over the square of the diffuson, Eq.~(\ref{st}),
is dominated by infrared divergences. In a 2-$d$ system, e.g.,
relevant energies extend to $\bar{\epsilon}\sim T$. For the following, it is convenient 
to rewrite Eq.~(\ref{tab}) [see also Eq.~(\ref{semi})] as 
 \begin{equation}
\frac{1}{\tp}=\frac{1}{\tau\sub{hit}} \frac{\langle \sigma_{\rm
    inel}\rangle}{\sigma_{\rm max}},
\end{equation}
where $\sigma_{\rm max}=4 \pi/k_F^2$ is the cross section of a unitary
scatterer, $\sigma_{\rm
    inel}/\sigma_{\rm max}$ is the conditional probability of {\em
    inelastic} scattering if an electron hits the impurity, and
 \begin{equation}
\frac{1}{\tau\sub{hit}}=\frac{2n_i}{\pi\nu}
\end{equation}
describes  the typical  'hitting rate'. 

As in the WL experiment discussed above, fits to experimental data have to be done with the $\epsilon$-independent dephasing rate. For a comparison with experiment we therefore have to give the $\epsilon$-independent dephasing rate which for $k=1$ is obtained by solving the equation

\begin{align}
\label{tp(tbl)}
&\frac{\Lp(T,B,L)}{L} e^{-\frac{L}{\Lp(T,B,L)}} \nonumber \\
&\hspace{1.2cm}=
\frac{3}{8T}\int d\epsilon \frac{e^{-\frac{L}{ \Lp(\epsilon,T,B) }}}{\cosh^4(\epsilon/2T)} \frac{\Lp(\epsilon,T,B)}{L},
\end{align} where $\Lp=\sqrt{D\tp}$ is the dephasing length. Notice that the actually measured dephasing rate depends on the length of the ring. It also differs from the WL result due to the different energy averages.

\section{Numerical results for dephasing rates from Aharanov-Bohm oscillations}
In order to evaluate the dependence of the dephasing rate on magnetic field, 
temperature, and ring length from Eq.(27), we require the T-matrix
for the single impurity Kondo model defined in Sec.~II. By using the 
equation of motion method this can be expressed as \cite{costi.00}
\begin{multline}
T_\sigma(\omega,T,B)=J \langle S_z \rangle +J^2
\left\langle\!\! \left\langle
 \bold{S} c^\dagger_\alpha {\boldsymbol \sigma}_{\alpha \sigma};
  \bold{S}  {\boldsymbol \sigma}_{\sigma \alpha'} c_{\alpha'} 
\right\rangle\!\!\right\rangle \label{eom}
\end{multline}
where $\langle\langle...\rangle\rangle$ denotes a retarded correlation
function and ${\boldsymbol \sigma}$ are the Pauli spin matrices. We calculate Eq.~(\ref{eom})
by applying the numerical renormalization group (NRG) method \cite{wilson.75+kww.80} 
for finite temperature dynamics \cite{costi.94}. At finite magnetic field, 
it is also important to use the reduced density matrix \cite{hofstetter.00} 
to evaluate the above dynamical quantity. For all calculations presented here
we used a discretization parameter for the conduction band of $\Lambda=1.5$ and
we retained 960 states per NRG iteration. We checked that this number of states
was sufficient to maintain particle-hole symmetry of the spectral densities 
$\opn{Im} T_{\uparrow}(\omega,T,B)= \opn{Im} T_{\downarrow}(-\omega,T,B)$ at this
relatively small value of the discretization parameter. The Friedel sum rule
for the $T=0$ spectral density was satisfied to more than 1\% accuracy in our
calculations.

\begin{figure}[h]
\vspace{.7cm}
\centering \includegraphics[width=8cm]{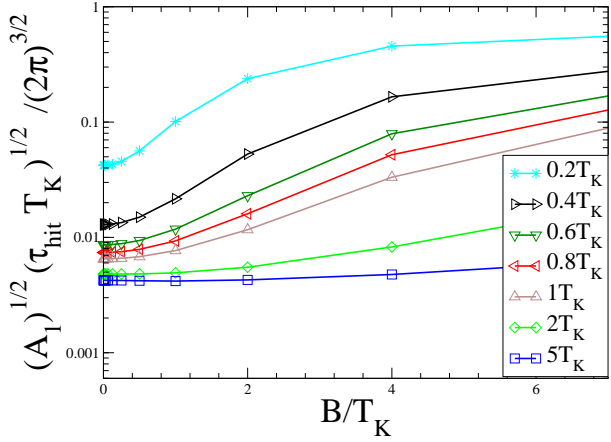}
\caption{\label{fig11} Amplitude of the Aharonov-Bohm oscillations [in units of
  $((2\pi)^3/(\tau\sub{hit}T\sub{K}))^{1/2}$] as
  a function of the applied magnetic field, $B$ (in units $T\sub{K}$,
  $\mu\sub{B}/k\sub{B}=1$), for different temperatures $T$ (in units
  $T\sub{K}$) obtained from NRG calculations for $L/L_{\rm hit}=10$.
  The rapid rise of the oscillation amplitude results from the
  suppression of dephasing (see Fig.~\ref{figa}) by polarizing the spins.
  Here, and in the remaining figures in this section, the symbols represent 
  the discrete values of (B,T) at which NRG calculations were carried out.
}
\end{figure}

Fig.~\ref{fig11} shows the numerical evaluation of $\sqrt{{\cal
    A}_1(B)}$ for various $T$ at a given length $L=10L\sub{hit}$ where
$L\sub{hit}=\sqrt{D \tau_{\rm hit}}$.  Notice that, if lengths are measured in
units of $L\sub{hit}$, the amplitude $(A_1 \tau_{\rm hit} T\sub{K})^{1/2}$
becomes a universal function of $B/T\sub{K}$ and $T/T\sub{K}$.

\begin{figure}[h]
\vspace{.7cm}
\centering \includegraphics[width=8cm]{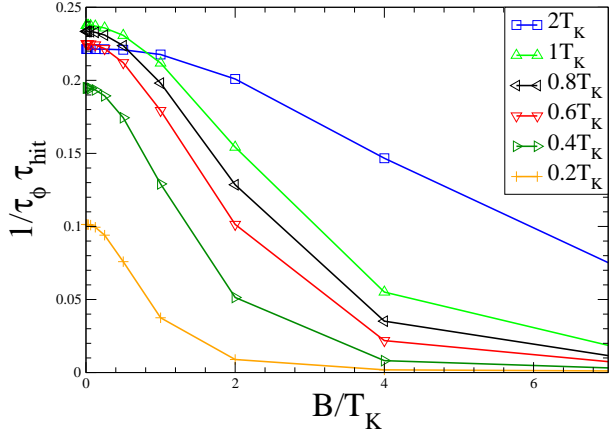}
\caption{\label{figa}The dephasing rate as a function of the magnetic
  field, $B$ (in units $T\sub{K}$) for various temperatures, $T$, and
  a ring of length $L=10L\sub{hit}$. Note the rapid suppression of
  $1/\tp$ especially for low temperatures. }
\end{figure}

Fig.~\ref{figa} gives the magnetic field dependence of the dephasing
rate at various $T$ and a fixed ring length $L=10L\sub{hit}$. For
large magnetic fields,  $B\gg T,T\sub{K}$, the dephasing rate is
expected\cite{VaGlLa03} to vanish proportional to
$(T/B)^2/\ln^4[B/T\sub{K}]$,
consistent with the numerical results (the precise form also depends
on $L/\Lp$, see below).

\begin{figure}[h]
\vspace{.7cm}
\centering \includegraphics[width=8cm]{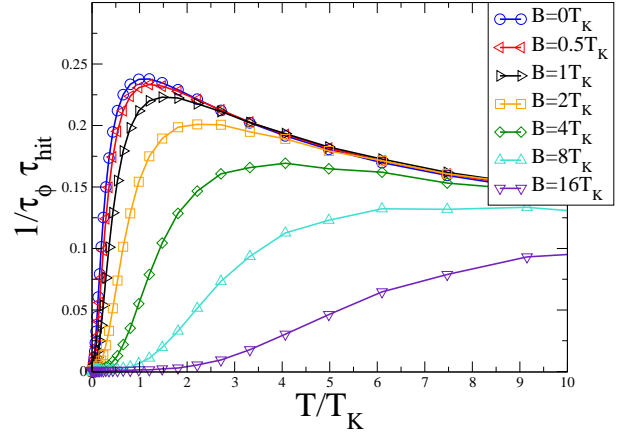}
\caption{\label{fig9} NRG results for the dephasing rate as a function
  of $T$ and for different values of
  $B$. $\tau\sub{hit}=\frac{\pi\nu}{2n\sub{i}}$ as defined above, $T$
  and $B$ are given in units of $T\sub{K}$~\cite{tk} (we set
  $\mu\sub{B}/k\sub{B}=1$). The values are for a ring of length, $L$,
  $L=10L\sub{hit}$ where $L\sub{hit}=\sqrt{D\tau\sub{hit}}$. While the
maximal dephasing occurs for $T\sim T\sub{K}$ for small fields $B\ll T_{K}$, 
it shifts to larger values ($T \sim B$) for $B \gg T\sub{K}$.}
\end{figure}

Fig.~\ref{fig9} shows the results for $1/\tps$ as a function of $T$
for different strengths of the magnetic field $B$ at a fixed ring
length $L=10L\sub{hit}$. While the
maximal dephasing occurs for $T\sim T\sub{K}$ for small fields $B\ll T\sub{K}$, it shifts
to larger values ($T \sim B$) for $B \gg T\sub{K}$. For high
temperatures, small fields and not too large $L/\Lp$, see below,
$T\gg T\sub{K}, B$ the dephasing rate is well described by the
Nagaoka-Suhl formula,~\cite{MiAlCoRo06}
$1/\tp(T)=\frac{n\sub{i}}{2\pi\nu}\frac{\pi^23/4}{\pi^23/4+\ln^2T/T\sub{K}}$.

\begin{figure}[b]
  \vspace{.7cm} \centering
  \includegraphics[width=0.95 \linewidth,clip]{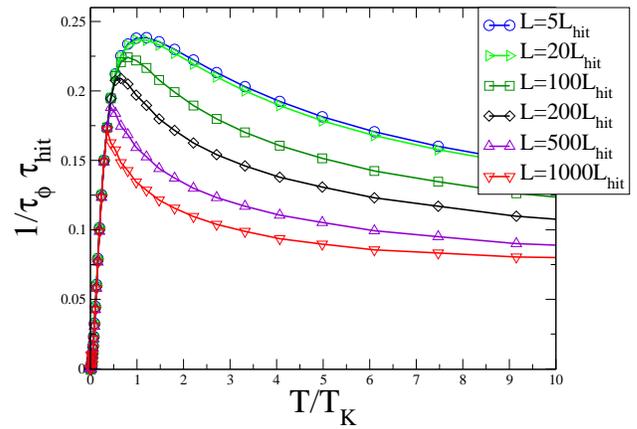}
\caption{\label{fig10} Temperature dependent dephasing rate for
  various ring lengths $L$, measured in units $L\sub{hit}=\sqrt{
    \frac{D\pi\nu}{2n\sub{i}}}$ and calculated via NRG for $B=0$. The
  logarithmic suppression with $L$ for larger $T$ arises due to the
  interference of electrons with energies larger than $T$. }
\end{figure}

Fig.~\ref{fig10} shows the dependence of $1/\tps(T)$ on the ring
length, $L$, at zero magnetic field.  As pointed out above, the
$L$-dependence of the experimentally measured dephasing rate enters
through the energy average of Eq.~(\ref{tp(tbl)}). As can be seen from
Fig.~\ref{fig10} the dephasing rate only changes by a factor $1/4$ on
increasing the ring length by a factor 200. We included curves for the
rather academic cases $L=100-1000L\sub{hit}$ (the amplitude is too
strongly suppressed to be observed for such large ring lengths) in
order to show this weak $L$-dependence. Fig.~\ref{fig10} also shows
that for temperatures $T\lesssim T\sub{K}$ the dephasing rate becomes
entirely $L$-independent. This reflects the fact that the energy
resolved dephasing rate, $1/\tps(\epsilon,T)$, establishes a deep
minimum at $\epsilon=0$ for $T\lesssim T\sub{K}$ and the integral on
the right hand side of Eq.~(\ref{tp(tbl)}) is therefore well
approximated by setting $\epsilon=0$. To be precise, in the limit of
ring lengths $L\gg \Lp$ the integral on the right hand side of
Eq.~(\ref{tp(tbl)}) (for fixed $T$ and $B$) is dominated by the saddle
points of the function

\begin{align*}
f(\epsilon)= \frac{L}{ \Lp^{(4)} (\epsilon) } - \ln\left[\frac{\Lp^{(4)}(\epsilon)}{L}\right] +4\ln\left[ \cosh\left(\frac{\epsilon}{2T}\right)\right].
\end{align*} For temperatures $T\lesssim \opn{max}\{B, T\sub{K}\}$ $f$
has a saddle point at $\epsilon=0$, which for temperatures $T\gtrsim
T\sub{K}$ becomes unstable. At very large ring diameters
$L/L\sub{hit}\gg 10^2$, a second saddle point at
$\epsilon=\frac{L}{2L\sub{hit}\ln^2\left[\frac{L}{2L\sub{hit}}\right]}T$
starts to dominate the integral for $T\gtrsim T\sub{K}$. Here
$L\sub{hit}=\sqrt{D\tau\sub{hit}}$ is the diffusive length scale
corresponding to the time $\tau\sub{hit}=\frac{\pi\nu}{2n\sub{i}}$ and
introduced above. Although this limit is rather academic it is
interesting that for such big rings dephasing is dominated by rare
events of highly excited thermal electrons scattering from the
magnetic impurities. This originates from the fact that high-energy electrons scatter less effectively from Kondo spins, as Kondo renormalization becomes less effective for $\epsilon \gg T\sub{K}$. Inserting this second saddle point into
Eq.~(\ref{tp(tbl)}) one finds that the length dependence of $1/\tp$
for high temperatures follows
\begin{align}
\frac{1}{\tp(T,L)} \sim \frac{1}{\ln^2\left[\frac{L}{2L\sub{hit}\ln^2\left[\frac{L}{2L\sub{hit}}\right]}\right]}
\end{align}
explaining the weak suppression of $1/\tp$ for large ring lengths shown
in Fig.~\ref{fig10}.

\section{Discussion and Conclusions}

In this paper we generalized previous results for the dephasing rate
due to diluted Kondo impurities as measured in the weak localization
experiment to describe dephasing due to {\em arbitrary} diluted
impurities.  Furthermore, we investigated how magnetic fields modify
the dephasing rate due to Kondo spins as can be measured 
 in mesoscopic  Aharonov-Bohm rings. 
We give results for the numerically evaluated dephasing rate as a
function of the magnetic field, temperature, and the ring length.

\begin{figure}[h]
\vspace{1cm}
\centering \includegraphics[width=1 \linewidth,clip]{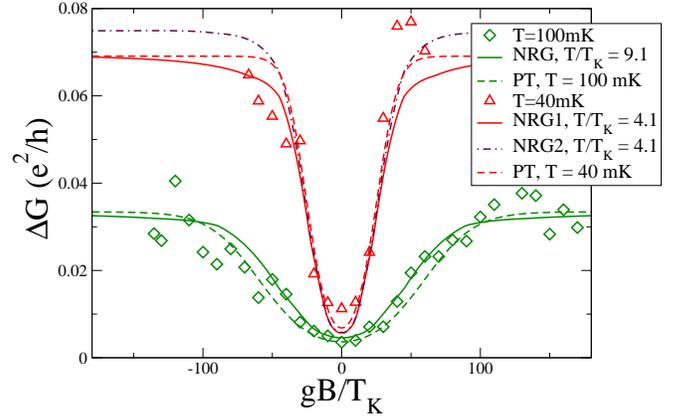}
\caption{\label{exp}Amplitude of the AB oscillations in units of
  $e^2/h$ as a function of $T/T_K$ for $T=40$\,mK $(\Delta)$ and
  $100$\,mK $(\diamond)$ measured by Piere {\it et al.} \cite{PieBir02,PiGoAnPoEsBi03},
  assuming $T_K=10$\,mK. Solid and dot-dashed lines are the numerically calculated
  amplitudes with fitting parameters described in the main text. As
  for very high magnetic fields, $B \gg 100\,T_K$, numerical errors
  increase when the dephasing rate becomes very small, we used an
  extrapolation of the numerical results, $1/\tp \propto 1/B^2$, in
  this regime. The saturation of the amplitude at these high fields
  arises as the dephasing due to electron-electron interaction
  dominates.  The data is equally well described by the fits used in
  Refs.~[\onlinecite{PieBir02}] and [\onlinecite{PiGoAnPoEsBi03}] (dashed lines), see main text. For the solid lines we used the same values for the dephasing rates  ($\tau_{ee}=5.4$\,ns and $9.9$\,ns for
$T=100$\,mK and $T=40$\,mK, respectively)  as in Ref.~[\onlinecite{PiGoAnPoEsBi03}], where $\tau_{ee}\propto T^{-2/3}$ was assumed. For the dot-dashed curve we use instead $\tau_{ee}=13.5$\,ns  for $T=40$\,mK since one expects theoretically\cite{tpab} that $\tau_{ee}\propto 1/T$ for $L \gg L_\phi$ (note, however, that $L \sim L_\phi$ in this regime explaining the rather weak dependence on $\tau_{ee}$).}
\end{figure}

The influence of magnetic impurities on dephasing has been studied in
a number of magneto-resistance experiments in Cu, Ag or Au wires doped
with magnetic impurities.~\cite{MoJaWe97,PiGoAnPoEsBi03,ScBaRaSa03,AlzBir06,MaBau06} More recently, high-precision
experiments using ion-implanted Fe impurities in Ag wires allowed a
quantitative comparison with our theory for spin-1/2 impurities, see
Refs.~[\onlinecite{MaBau06}] and [\onlinecite{AlzBir06}] for a critical discussion.
Such studies using samples doped with magnetic impurities have, to our knowledge, 
only been performed  in the spin-glass regime \cite{spinglassregime} using 
magnetic ions with tiny Kondo temperatures.   Pierre {\it et  al.} \cite{PieBir02,PiGoAnPoEsBi03} studied rings made from nominally clean Cu wires.
As these wires show a saturation of the dephasing rate (determined from
weak localization) at low temperatures ($30$\,mK$\lesssim T \lesssim
1$\,K), it was suspected that tiny concentrations of magnetic
impurities with Kondo temperatures below 30\,mK may be at the origin
of the observed saturation.  This picture could be confirmed as
measurements of the amplitude of Aharonov-Bohm oscillation displayed a
dramatic rise by almost an order of magnitude in moderate magnetic
fields (see Fig.~\ref{exp}), proving the magnetic origin of the
low-$B$, low-$T$ dephasing.

As neither the concentrations nor the type(s) of magnetic impurities
are known, a parameter-free comparison to our predictions is not
possible for these systems. Assuming Mn impurities, believed to be
characterized by a Kondo temperature of the order of 10\,mK,~\cite{wohlleben} and, using
the same dephasing rates due to electron-electron interactions as in
Ref.~[\onlinecite{PiGoAnPoEsBi03}] ($\tau_{ee}=5.4$\,ns and $9.9$\,ns for
$T=100$\,mK and $T=40$\,mK, respectively) we obtain the fits shown in
Fig.~\ref{exp} for a $g$ factor of $g\approx 1.4$ and an impurity
concentration of 2.7\,ppm. We have also added a curve at $T=40$\,mK (dot-dashed line) which uses 
$\tau_{ee}=13.5$\,ns (keeping all other parameters identical) to take into account that one expects theoretically\cite{tpab} $\tau_{ee}\propto 1/T$.  The fits and the extracted parameters are
not very reliable as can be seen from the observation that the data
has been equally well described in Refs.~[\onlinecite{PiGoAnPoEsBi03}] and [\onlinecite{PieBir02}] by the
simple perturbative formula $\frac{\tps(B=0)}{\tps(B)}
=\frac{g\mu_B/k_BT}{\sinh(g\mu_B/k_BT)}$ with $g=1.08$, see Fig.~\ref{exp}.

For a more meaningful comparison to our results, 
experiments on AB rings, doped with magnetic impurities with a higher
Kondo temperature would be highly desirable. On the one hand, such experiments and their theoretical interpretation can reveal basic dephasing mechanisms in metals, on the other hand, they can be used to obtain insight into the physics of strongly correlated dynamical impurities and their interactions.

We thank
 Ch.~B\"auerle, N. Birge, J.~v.~Delft, L.~Glazman, S.~Kettemann, S.~Mirlin, J. Mydosh, L. Saminadayar,
 B.~Spivak, P.~W\"olfle, and, especially, A. Altland for useful discussions and  N.~Birge for sending us his experimental data. Furthermore we acknowledge financial support from the 
Deutsche Forschungsgemeinschaft through the SFB 608 
and Transregio SFB 12 and the NIC Juelich for computing time

\bibliography{tobias}

\begin{thebibliography}{}
\bibitem{h2e}
The $h/2e$-AB oscillations arise by the same mechanism as the WL and are suppressed when the magnetic field penetrates the wire of the sample.
\bibitem{mirlinring}
Notice that in the AB experiment (considered below)
the suppression of inelastic vertex contributions is even larger and proportional to 
$1/(\Delta E t\sub{D})^2\ll 1$ (Ref.~[\onlinecite{tpab}]), where $t\sub{D}=L^2/D\gg \tp$ and $L$ is the ring length. This is due to the fact that  in this case the typical time scale of interfering electrons is $t\sub{D}$ rather than $\tp$ as electrons contributing the AB oscillations have to circle the ring at least once.
\bibitem{tk}
We define and determine the Kondo temperature $T\sub{K}$ from the $T=0$ suceptiblity $\chi=(g\mu\sub{B})^2/(4T\sub{K})$.
\end{thebibliography}

\bibliographystyle{prsty}

\pagestyle{empty}

\end{document}